\documentclass[prl,twocolumn,showpacs,preprintnumbers,amsmath,amssymb]{revtex4}

\usepackage{graphicx}
\usepackage{amsmath}
\usepackage{amsbsy}
\usepackage{dcolumn}
\usepackage{bm}
\begin{document}

\title{
Spin-polarized electronic structures and transport properties of Fe-Co alloys
}

\author{Yohei Kota}
\affiliation{Department of Applied Physics, Tohoku University \\
6-6-05, Aoba, Aoba-ku, Sendai 980-8579, Japan}
\author{Tomohiro Takahashi}
\affiliation{Department of Applied Physics, Tohoku University \\
6-6-05, Aoba, Aoba-ku, Sendai 980-8579, Japan}
\author{Hiroki Tsuchiura}
\affiliation{Department of Applied Physics, Tohoku University \\
6-6-05, Aoba, Aoba-ku, Sendai 980-8579, Japan}
\author{Akimasa Sakuma}
\affiliation{Department of Applied Physics, Tohoku University \\
6-6-05, Aoba, Aoba-ku, Sendai 980-8579, Japan}

\date{\today}

\begin{abstract}

The electrical resistivities of Fe-Co alloys owing to random alloy disorder are calculated
using the Kubo-Greenwood formula.
The obtained electrical resistivities agree well with experimental data quantitatively at low temperature.
The spin-polarization of Fe$_{50}$Co$_{50}$ estimated from the conductivity (86\%)
has opposite sign to that from the densities of the states at the Fermi level (-73\%).
It is found that the conductivity is governed mainly by s-electrons, and the s-electrons
in the minority spin states are less conductive due to strong scattering by the large densities of the states
of d-electrons than the majority spin electrons.

\end{abstract}


\maketitle



Fe-Co alloys are used for soft magnetic materials because they have low magnetic anisotropy and
relatively large magnetic moment per atom among various binary alloys of 3d transition metals.
These alloys are promising materials for ferromagnetic layers in the giant magnetoresistance (GMR)
and the tunnel magnetoresistance (TMR) devices which are applied for the magnetoresistive heads and so on.
Since the performances of such devices are strongly affected by the transport properties of magnetic materials,
the spin-dependent transport properties are needed to study in the view of the quantum transport theory.



The Slater-Pauling curve of Fe-Co alloys has the peak when the Co concentration is about 30\% \cite{ref01}.
These alloys have the bcc-structure when the Co concentration is less than 75\%, and are transformed
into the fcc or hcp-structures more than 75\% \cite{ref02}.
There are a lot of works on the electronic structures of Fe-Co alloys based on ab-initio calculations,
for instance Refs. 3, and they are in fairly good agreement with experimental results.
In the present work, an ab-inito calculation is performed for the electronic structures and conductivities 
of Fe$_{100-x}$Co$_x$ ($x=10,20,...,70$) bcc-based disordered alloys using a tight-binding (TB)
linear muffin-tin orbital (LMTO) method based on the local spin-density approximation (LSDA)
combined with the coherent potential approximation (CPA) \cite{ref04}.
The conductivities of these systems are estimated based on the Kubo-Greenwood formula \cite{ref05, ref06}
within the framework of the linear response theory.
We employ the method developed by Kudrnovsk\'y and his co-workers for the actual calculation
\cite{ref04, ref07, ref08}.
To calculate the Green's functions, we took $\delta=0.5$ mRy as the infinitesimal damping constant $\delta$.
In this case, 10$^5$ k-points are needed to achieve the sufficient accuracy.



\begin{figure}
\includegraphics[clip,width=7cm]{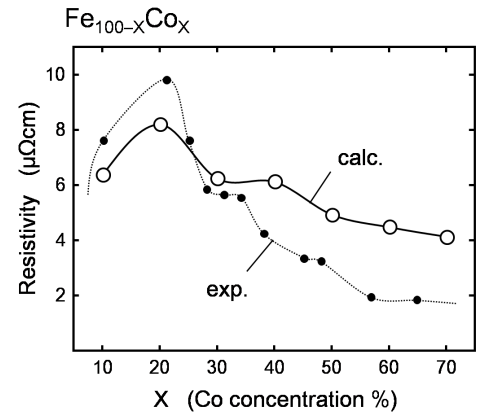}
\caption{
Electrical resistivities of Fe$_{100-x}$Co$_x$ alloys as functions of Co concentration $x$.
The open circles indicate the calculated results and closed circles experimental data.
}
\label{fig1}
\end{figure}

Figure \ref{fig1} shows the calculated results and experimental data of the bulk Fe-Co alloys
at low temperature (4.2 K) \cite{ref09} as functions of Co concentration $x$.
In the experimental curve of this data, the resistivity of Fe-Co alloys exhibits a peak around $x=20$.
The calculated results show quantitative correspondence with this data.
Note, however, that the additional resistivity due to $\delta$ is found to be around 2.0 $\mu \Omega cm$.
Thus the precious results are considerably smaller than the experiments in the whole range of $x$,
which is rather natural since in the present work the effects such as impurities, defects and so on are not taken into account.

\begin{figure}
\includegraphics[clip,width=7cm]{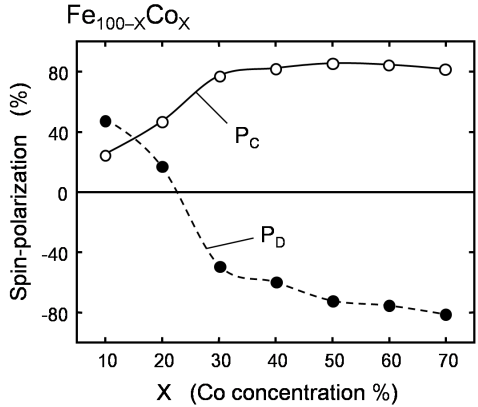}
\caption{
Spin-polarization of the DOS at $E_F$ (closed circles) and the conductivities (open circles)
of Fe$_{100-x}$Co$_x$ alloys as functions of Co concentration $x$ respectively.
}
\label{fig2}
\end{figure}

We next show the spin-polarizations estimated from the densities of the states (DOS) at the Fermi level ($E_F$)
and the electrical conductivities as functions of Co concentration $x$ in Fig. \ref{fig2}.
With increasing $x$, the sign of the spin-polarization estimated from the DOS ($P_D$) changes from positive
to negative value around $x=25$ and the value increases for $x>25$, while the spin-polarization
estimated from the conductivities ($P_C$) exhibits positive sign in the whole region of $x$.
It seems that $P_D$ and $P_C$ show opposite dependence on $x$.
In fact, in the region of $x>40$ the values of $P_D$ are still less than -50\%, while $P_C$ are over 80\%.
These results are inconsistent with the Boltzmann transport phenomenalism that conductivity is proportional to DOS,
and these imply that the usual way to estimate the spin-polarization such as the magnetoresistive ratio
of GMR and TMR devises from $P_D$ needs to be reconsidered.
The results remind us that most electrons at $E_F$ do not necessarily behave equally as conductive particles,
because of the difference of the mobility depending on the state electrons belong to.
Although most of the DOS at $E_F$ are occupied by heavy d-electrons, the electrical conduction is dominated
by light s-electrons.
Therefore we have to discuss the spin-dependent transport phenomena by considering both DOS and conductivity
which depend strongly on the electronic band structures.

\begin{figure}
\includegraphics[clip,width=8cm]{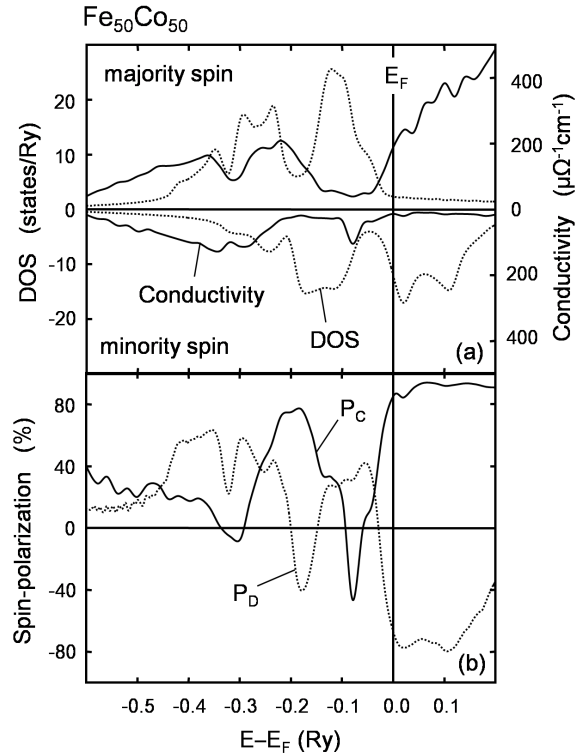}
\caption{
(a) Spin-dependent DOS (dashed lines) and conductivities (solid lines) of Fe$_{50}$Co$_{50}$ alloy
as a function of one-electron energy $E-E_F$ respectively.
(b) Spin-polarizations of DOS (dashed line) and conductivities (solid line).
}
\label{fig3}
\end{figure}

In order to investigate the relationship between the electronic structure and transport property of a ferromagnet,
we show the energy dependence of the spin-dependent conductivities evaluated based upon the rigid band scheme
together with the DOS of Fe$_{50}$Co$_{50}$ alloy in Fig. \ref{fig3} (a) and their spin-polarizations
in Fig. \ref{fig3} (b).
Fe$_{50}$Co$_{50}$ is highly favorable material in the field of spin-electronics and actively applied
for the ferromagnetic layer of magnetoresistive head.
We can understand, from the behavior of the spin-dependent conductivities, that the decrease of the $P_C$
in the low Co concentration region in Fig. \ref{fig2} is associated with the rapid decrease of the conductivity
of the majority spin electrons below $E_F$ as shown in Fig. \ref{fig3} (a).
Furthermore, it can be considered that much different feature of the energy dependence of the DOS and the conductivities
in Fig. \ref{fig3} (a) are closely related to the opposite behavior in Fig. 3 (b) as well as the Co concentration dependence
in Fig. \ref{fig2}.
Particularly, paying attention to the energy region above $E_F$, the conductivities
of the majority spin electrons grow steeply in spite of their small DOS's, while those of the minority spin electrons
having large DOS exhibit much smaller values than those of the majority spin electrons.
As a result, $P_C$ reaches closely to 90\% when one-electron energy shifts to higher side.
Therefore, it is naturally supposed that, in the present case, the d-electrons dominating most part
of the DOS make few contributions to the electrical conduction, whereas the s-electrons contribute much.
However, there is also no remarkable relationship between the DOS of s-states and the conductivities, so the main factor
for the conductivities other than the DOS of s-states must be considered.

Generally, the dominant factors of electrical conductivity are the velocity and the lifetime of electrons.
In the present work, the lifetime can be easily estimated from the inverse of the imaginary part
of the coherent potential functions $\{ Im \ \bm{P}\}^{-1}$  which is proportional
to the imaginary part of the self-energy describing scattering probability due to random alloy disorder.
Figure 4 shows the inverse of the imaginary part of the coherent potential functions $\{ Im \ \bm{P}\}^{-1}$
of d-electrons for each spin state and the spin-polarization ($P_{\tau}$) estimated from these values
as a function of one-electron energy relative to $E_F$.
The $P_{\tau}$ of d-electrons exhibits abrupt increase just below $E_F$ and keeps high level,
which implies that the scattering probability of the minority spin electrons is much higher than that
of the majority spin electrons at $E_F$.
Looking at the energy dependence of $P_C$ again in Fig. \ref{fig3} (b), one can notice that the behavior of $P_C$
is closely similar to that of the $P_{\tau}$ of d-electrons.
We have confirmed that the $P_{\tau}$ of s-electrons are much smaller than that of d-electrons and behave
as almost flat curve, much different from the $P_{\tau}$ of d-electrons.
From above results, we consider that the s-electrons having main contribution to the conduction
undergo strong hybridization with d-states and then the lifetime not only of d-electrons
but also of s-electrons are governed mainly by the scattering probability of d-electrons.
This leads to the situation that $P_C$ exhibits similar behavior to the $P_{\tau}$ of d-electrons
even though the major carrier is s-electrons.
Therefore s-electrons with high mobility are much influenced by this scattering factor.
In other words, the large DOS of d-states in the minority spin states have an effect to disturb the conduction
of s-electrons through s-d scattering, while the s-electrons in the majority spin states can conduct
with less scattering effect.
This is one of the reasons that the spin-polarization of the conductivities becomes too large.

\begin{figure}
\includegraphics[clip,width=8cm]{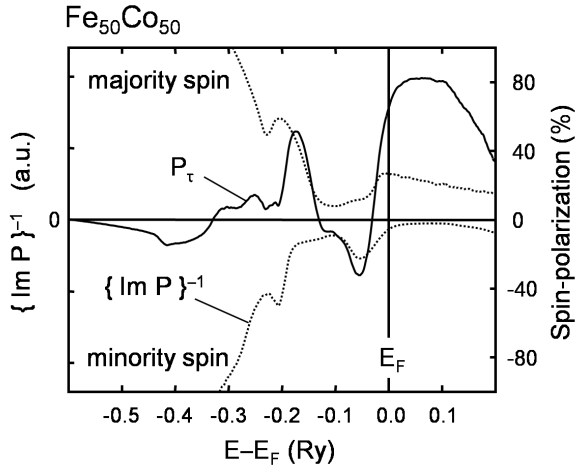}
\caption{
The inverse of the imaginary part of the coherent potential functions $\{ Im \ \bm{P}\}^{-1}$ on the d-states
of Fe$_{50}$Co$_{50}$ alloy for majority and minority spin (dashed lines) and the degree of their spin-polarization
(solid line) as a function of one-electron energy $E-E_F$ respectively.
}
\label{fig4}
\end{figure}

The spin-polarization of Fe$_{50}$Co$_{50}$ reaches 86\% in the present work and this result is close to
the experimental estimation from the measurement of the current perpendicular to plane magnetoresistance
(CPP-MR) analyzed by the two current model \cite{ref10,ref11}.
It is reported that the spin-polarization of CoFe becomes about 85$\pm$10\%.
Theoretically, the spin-polarizations of 3d transition metals were estimated by Bahramy, {\itshape et al.} \cite{ref12}
and Zhu, {\itshape et al.} \cite{ref13} based on the equation formulated by Mazin \cite{ref14}.
The average of the group velocities estimated from the spin-dependent dispersions of the majority spin states is larger
than that of the minority spin in their works and they obtained the spin-polarizations of positive sign as well as our result.



In this work, TB-LMTO combined with CPA is performed to calculate the electronic structures on Fe-Co systems.
The electrical conductivities are calculated based on the Kubo-Greenwood formula and the technique developed
by Kudrnovsk\'y and his co-workers.
The results of the electrical resistivity are quantitatively agreeable with the experimental data at low temperature.
The spin-polarizations estimated from the DOS at $E_F$ and from the conductivities exhibit quite different behavior
as functions of Co concentration, because conductivity is mainly dominated by s-electrons although DOS
are almost occupied by d-electrons.
In the case of Fe$_{50}$Co$_{50}$, the s-electrons having majority spin can conduct with less scattering processes,
while those in the minority spin states undergo strong s-d scattering due to the large DOS and the short lifetime
of d-electrons.
This causes a remarkable difference between the spin-polarizations of the DOS at $E_F$ (-73\%)
and the conductivity (86\%).

{\it Acknowledgments.} 
Numerical computation in this work was partially carried out by using supercomputing resources
at Information Synergy Center, Tohoku University and the Yukawa Institute Computer Facility.
This work is supported by Grant-in-Aids from ``Education Program for Biomedical and Nano-Electronics,
Tohoku University'' Support Program for Improving Graduate School Education, MEXT,
and from the Ministry of Education, Culture, Sport, Science and Technology of Japan.


\newpage

\end{document}